\documentclass[aps,prl,showpacs,twocolumn,superscriptaddress,amsmath,amssymb,floatfix]{revtex4}
\usepackage{tabularx}
\usepackage{bm}
\usepackage{epsfig,subfigure}
\usepackage{multirow}
\usepackage{graphicx}
\usepackage{color}
\usepackage{amsfonts}
\usepackage{exscale}
\usepackage{amsbsy}
\graphicspath{{Fig/}}
\usepackage[colorlinks,urlcolor=blue,linkcolor=blue,anchorcolor=blue,citecolor=blue]{hyperref}
\newcommand{\tabincell}[2]{\begin{tabular}{@{}#1@{}}#2\end{tabular}}
\newcommand{\mr}{\multirow}

\begin{document}
\title{Bosonic Integer Quantum Hall effect in an interacting lattice model}
\author{Yin-Chen He}
\affiliation{Max-Planck-Institut f\"{u}r Physik komplexer Systeme, N\"{o}thnitzer Str. 38, 01187 Dresden, Germany}
\author{ Subhro Bhattacharjee} 
\affiliation{Max-Planck-Institut f\"{u}r Physik komplexer Systeme, N\"{o}thnitzer Str. 38, 01187 Dresden, Germany}
\affiliation{International Centre for Theoretical Sciences, Tata Institute of Fundamental Research, Bangalore 560012, India}
\author{R. Moessner}
\affiliation{Max-Planck-Institut f\"{u}r Physik komplexer Systeme, N\"{o}thnitzer Str. 38, 01187 Dresden, Germany}
\author{Frank Pollmann}
\affiliation{Max-Planck-Institut f\"{u}r Physik komplexer Systeme, N\"{o}thnitzer Str. 38, 01187 Dresden, Germany}
\begin{abstract}
We study a bosonic model with correlated hopping on a honeycomb lattice, and show that its ground state is a bosonic integer quantum Hall (BIQH) phase, a prominent example of a symmetry protected topological (SPT) phase. By using the infinite density matrix renormalization group method, we establish the existence of the BIQH phase by providing clear numerical evidence: (i) a quantized Hall conductance with $|\sigma_{xy}|= 2$ (ii) two counter propagating gapless edge modes. Our simple model is an example of a novel class of systems that can stabilize SPT phases protected by a continuous symmetry on lattices and opens up new possibilities for the experimental realization of these exotic phases.
\end{abstract}
\pacs{73.43.-f, 03.65.Vf }
\maketitle

\textbf{Introduction:} Topological phases, which are quantum states of matter beyond Landau's symmetry breaking paradigm, are a main focus of  modern condensed matter physics \cite{Wen_book}. 
Well known examples are fractional quantum Hall (FQH) phases in two-dimensional electron systems \cite{QHE} and quantum spin liquids that can arise in frustrated spin systems \cite{Anderson1973,Moessner2001, Wen_book}. 
These examples fall into the category of {intrinsic  topological order} \cite{Wen1990},  characterized by their fractionalized quasiparticle excitations and the presence of {long-range entanglement}. 
More recently it was found that phases with only {short range entanglement} can also be topological non-trivial if certain symmetry is obeyed \cite{Gu2009,Pollmann2010,Chen2012, Chen2013}. 
Such {symmetry protected topological} (SPT) phases have a gapped bulk with no fractionalization and are usually characterized by their  anomalous  gapless \cite{Chen2011,Levin2012} or intrinsic topologically ordered \cite{Vishwanath2013,Metlitski2013,Burnell2014,Ye2015} edges.  

SPT phases in $d$-dimensional systems, protected by a global symmetry $G$, are generally classified by the cohomology group $H^{d+1}[G, U(1)]$  \cite{Chen2012, Chen2013}.
The first example discussed in this context was the Haldane spin chain \cite{Haldane1983,Affleck1987}, which is a one-dimensional SPT phase protected by either spin rotational (e.g., $\mathbb{Z}_2\times\mathbb{Z}_2$), time reversal or inversion symmetry \cite{,Pollmann2010}. 
The SPT phases protected by continuous symmetries can be described using effective topological field theories \cite{Vishwanath2013, Lu2012,Xu2013,Senthil2013,Liu2013}. 
For example, the bosonic integer quantum Hall (BIQH) state,  which is a $U(1)$ SPT phase (protected by the $U(1)$ charge conservation), can be described by a mutual Chern-Simons theory \cite{Lu2012,Senthil2013}. 
Along this line, BIQH states can be constructed using composite bosons in quantum Hall systems \cite{Senthil2013} and were found to be stabilized in  two-component bosonic continuum models in a strong orbital magnetic field \cite{Furukawa2013, Wu2013, Regnault2013}.

Bosonic SPT phases require interactions and have no simple non-interacting analogs, making it challenging to study and stabilize these exotic phases.   
Recently, several fine tuned exactly solvable lattice models have been introduced that realize  SPT phases that are protected by discrete symmetries \cite{Chen2011, Levin2012,Burnell2014}. 
In contrast, the realization of more physical SPT phases, which are protected by continuous symmetries, has rarely been explored \cite{Geraedts2013, Liu2014, Wang2015}.
It is  desirable to have  concrete lattice models as they allow for a  relatively simple numerical study of topological phase transitions \cite{Grover2013, Lu2014,Barkeshli2013} and  serve as a guide to where these  phases might appear in nature, either in strongly correlated systems or optical lattices.
Interestingly, an early study \cite{Moeller09} (before the concept of  SPT was introduced), has found a possible BIQH state  in a Hofstadter type model by calculating the Hall coefficients $\sigma_{xy} = 2$  using exact diagonalization on small system size. 
However, it is still unclear whether a $U(1)$ SPT phase is realized in such model, since  the existence of anomalous gapless edges, the most defining property of the  SPT phases, has not been verified.

\begin{figure}
\includegraphics[width=0.45\textwidth]{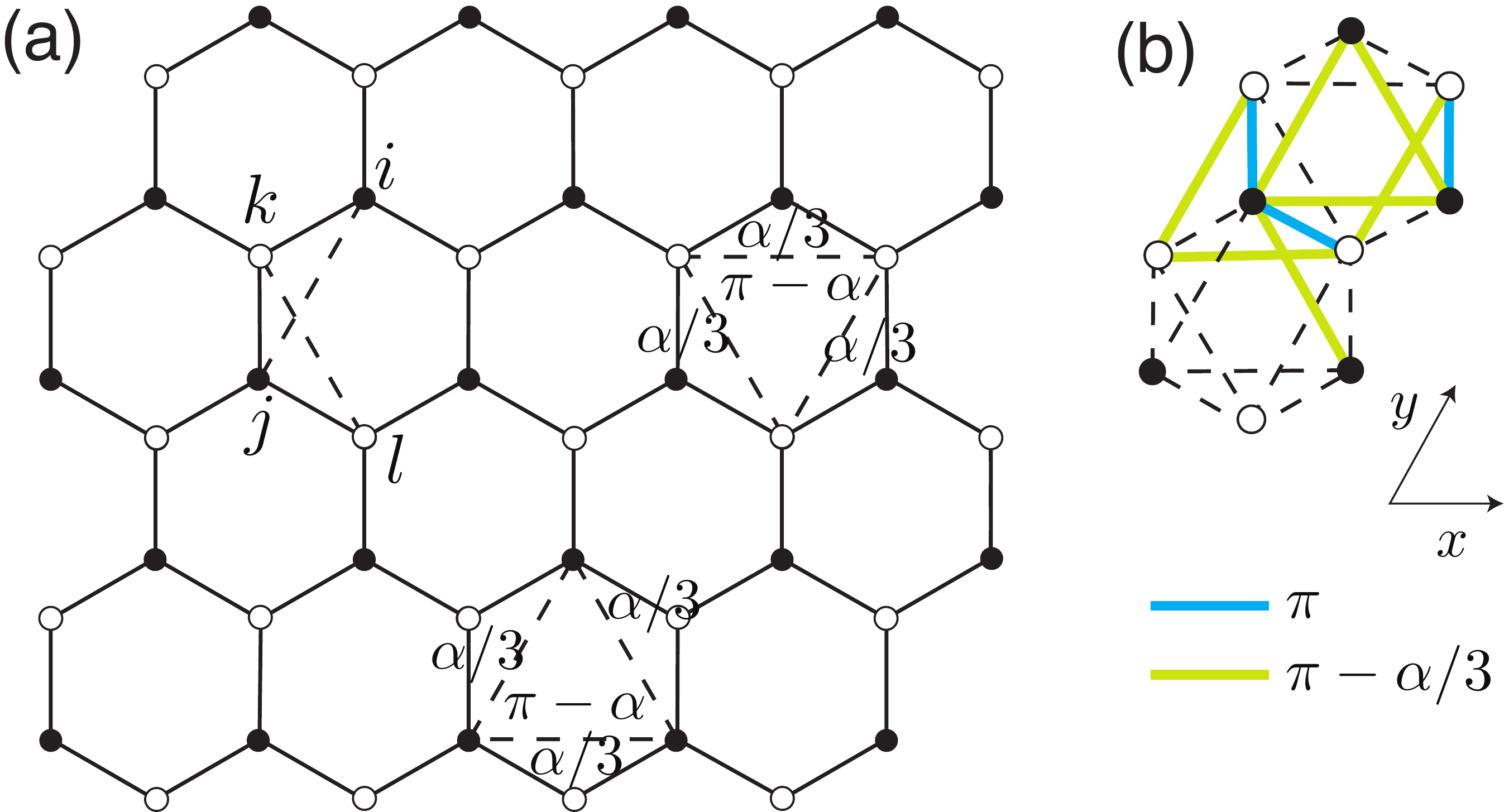} \caption{\label{fig:model} (color online). The model is defined on a honeycomb lattice, and we assign time-reversal symmetry breaking background flux in the system. (b) The unit cell is doubled to $4$ sites (along the $y$ direction). The gauge field on the specified link  is taken as $\pi$, $\pi-\alpha/3$. For the dashed links, the gauge fields are $0$ (nearest neighbors), $\alpha/3$ (next-nearest neighbors).}
\end{figure}

In this letter, we introduce  a simple lattice model for which we show numerically that it  realizes a BIQH phase.
Our model consists of hard-core bosons on a honeycomb lattice with correlated hopping which is subject to a background gauge flux, and hence breaks time-reversal symmetry explicitly. 
By using the infinite DMRG method \cite{McCulloch2008,DMRG} on a cylinder, we obtain numerical evidence that the model stabilizes two different BIQH phases with opposite quantized Hall conductances $\sigma_{xy}=\pm 2$, as the background flux is tuned. 
Firstly, when adiabatically inserting $2\pi$ flux, we find that two bosons are pumped from the left edge of cylinder to the right edge, hence the ground state has a quantized Hall conductance $\sigma_{xy}=2$ \cite{Laughlin1981}. 
Secondly, using entanglement spectra as a probe \cite{Li2008}, we show that the ground state has two counter-propagating edge modes, and it fits  theoretical expectation perfectly.

\textbf{Model: }
Our model is defined on a honeycomb lattice with hard-core bosons at half filling described by  the Hamiltonian
\begin{align}
H&=\sum_{\langle \langle ij \rangle\rangle} \left[e^{i \mathcal  A_{ij}} (2 n_k^b-1) a^\dag_i a_j +h.c.\right] \nonumber \\ &+ \sum_{\langle \langle kl \rangle\rangle} \left[e^{i\mathcal A_{kl}}(2 n_j^a-1) b^\dag_k b_l + h.c.\right]
\nonumber \\ &+\lambda \sum_{\langle k j\rangle} (e^{i\mathcal A_{kj}}a^\dag_k b_j+h.c. ), \label{eq:BIQH_Ham}
\end{align}
where $a$ ($b$) is the annihilation operator for a hard-core boson on sublattice $A$ ($B$) of the honeycomb lattice (see Fig. \ref{fig:model}). 
The first two terms are the correlated hopping terms: the hopping  of bosons on the sublattice $A$ are coupled to the particle number of bosons on the sublattice $B$ on the intermediate site, $n_k^b=b^\dag_k b_k=0, 1$; and vice versa.  
The last term is a standard nearest-neighbor hopping term.
We also assign a background flux $\phi_{\mathcal A}$ (Fig. \ref{fig:model}(a)): the triangular plaquette with three vertices from the same sublattice has a flux $\pi-\alpha$, the small triangular plaquette with vertices from different sublattices has a flux $\alpha/3$. Thus each hexagonal plaquette has a flux $\pi$.

When $\lambda=0$, the model has a $U(1)\times U(1)$ symmetry, which can  be considered as either the particle conservation on each sublattice, or the overall particle (charge) $n_a+n_b$ conservation plus the (pseudo) spin $n_a -n_b$ conservation.   
However, the BIQH state only requires one $U(1)$ symmetry to protect it.  
That is to say, we can  allow tunneling between sublattices ($\lambda\ne 0$),  so that the $U(1)\times U(1)$ symmetry is broken down to a global $U(1)$ from overall charge conservation, without destroying the BIQH phase. 
We find that this is in fact true for our model: the BIQH state is robust as we turn on a finite $\lambda$ ($\le 0.7$), 
and almost all the properties are the same as the $\lambda=0$ case.

We use the infinite DMRG method  \cite{DMRG, McCulloch2008} to study the system wrapped around a cylinder with two different geometries (see supplementary materials) and find that the results do not depend on this choice.  
In the following, we set for simplicity $\lambda=0$ (unless specified otherwise).  
Due to the background  flux  $\pi$ (in each hexagon) we double the unit cell to $4$ sites and we choose a gauge as shown in Fig. \ref{fig:model}(b). 
Note that it is possible that certain choices of gauge and geometry will give rise to an extra flux for a non-contractible loop along the $y$ direction (which can only be eliminated by a large gauge transformation).  
As the BIQH phase is gapped, such a global gauge flux will not cause any significant effect for our purpose. 

Based on the numerical simulations on systems of width  $W_y=8, 12, 16$ sites (corresponding to $L_y=2,3,4$ unit cells), we have obtained the phase diagram consisting two BIQH states with  opposite Hall conductance $\sigma_{xy}$: for $\alpha\in (0,\pi)$, $\sigma_{xy}=2$; for $\alpha \in (-\pi, 0)$, $\sigma_{xy}=-2$. %
To establish the existence of the BIQH phases,  we now discuss in detail two of their characteristic fingerprints: (i) the ground state has a quantized Hall conductance $|\sigma_{xy}|=2$ ;  (ii) the ground state has two counter-propagating gapless edge modes.

\begin{figure}
\includegraphics[width=0.45\textwidth]{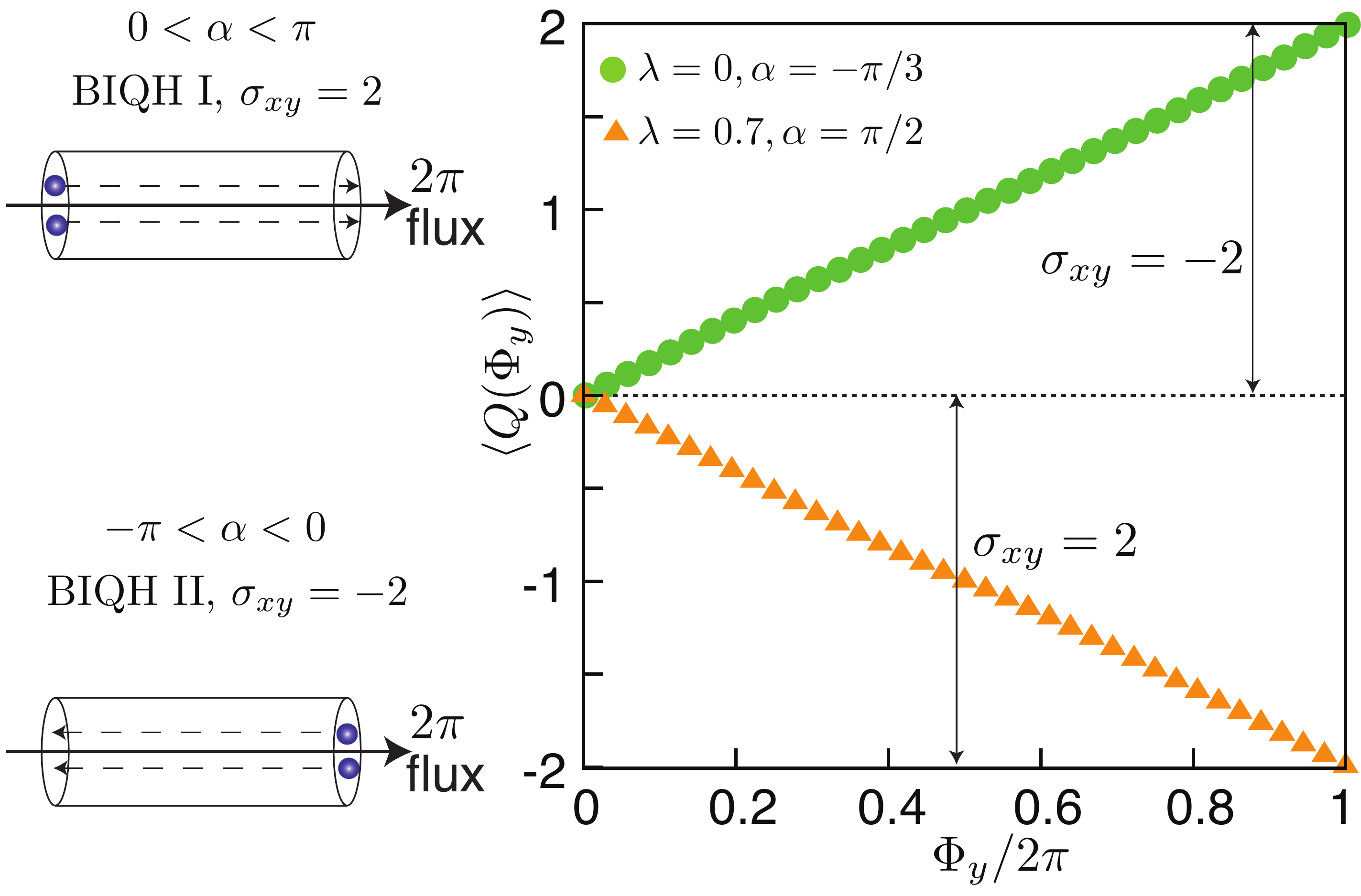} \caption{\label{fig:charge_pumping} (color online). Charge pumping in the BIQH state.  Here we show two cases (the  width is $W_y=8$): $\lambda=0.7$, $\alpha=\pi/2$, with Hall conductance $\sigma_{xy}=2$; $\lambda=0$, $\alpha=-\pi/3$,  with Hall conductance $\sigma_{xy}=-2$. For other system size, the results are almost the same.}
\end{figure}

\textbf{Quantized Hall conductance:} The quantized Hall conductance is a hallmark of the quantum Hall state. 
In contrast to fermionic systems, the Hall conductance $\sigma_{xy}$ of a BIQH state is always quantized to an even number \cite{Senthil2013}. 
Numerically, we can use an adiabatic flux insertion to measure the Hall conductance $\sigma_{xy}$: $2\pi$ flux insertion on a cylinder will pump $\sigma_{xy}$ particles from the left edge to the right edge of cylinder \cite{Laughlin1981}. 
Such flux insertion can be implemented in the Hamiltonian by twisting the boundary condition in the infinite DMRG algorithm \cite{ He2014}: the bosons hopping around the cylinder pick up a flux $\Phi_y$.  
The Hall conductance can then be written as \cite{Hall_conductance}:
\begin{align}
\sigma_{xy}&=\int_{0}^{2\pi}  [\partial_{\Phi_y} \langle Q(\Phi_y)\rangle] d \Phi_y, \\
 \langle Q(\Phi_y)\rangle &= \sum_{i} \lambda_i(\Phi_y) Q_i(\Phi_y),
\end{align}
where $\lambda_i(\Phi_y)$ are the eigenvalues of the reduced density matrix when flux $\Phi_y$ is inserted, $Q_i(\Phi_y)$ is the corresponding $U(1)$ quantum number.
The numerical data in Fig. \ref{fig:charge_pumping} clearly shows a quantized Hall conductance as $2\pi$ flux is inserted, and it is robust when turning on a finite $\lambda$. 

\begin{figure}
\includegraphics[width=0.45\textwidth]{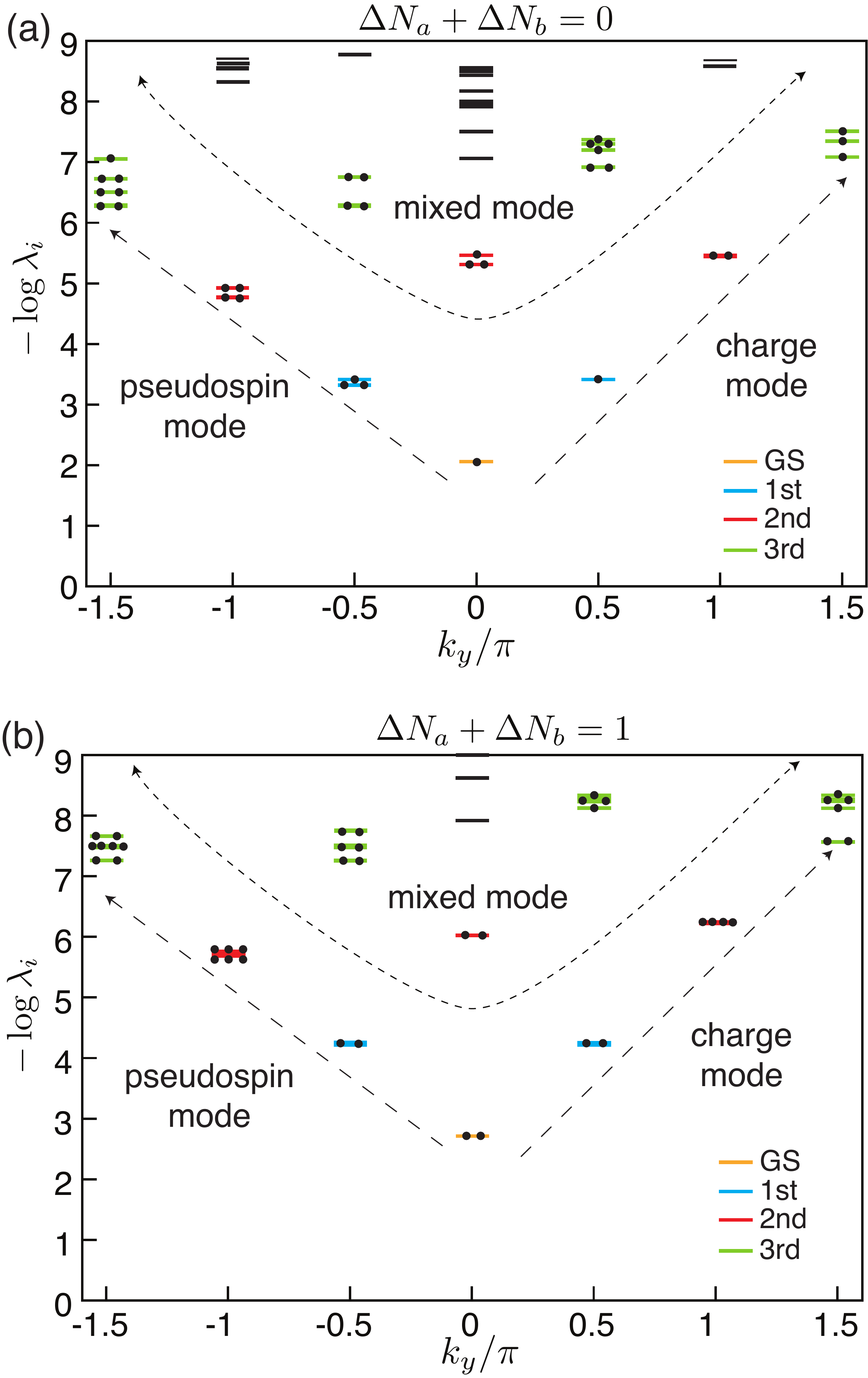} \caption{\label{fig:ES} (color online). The entanglement spectra versus momentum $k_y$: (a) charge sector $\Delta N_a+\Delta N_b=0$. (b) charge sector $\Delta N_a+\Delta N_b=1$. The simulation is carried on an  infinite cylinder of width $L_y=4$ unit cells ($W_y=16$ sites), $\alpha=\pi/2$, $\lambda=0$. The dashed arrow-lines denote the edge modes.}
\end{figure}

\textbf{Edge modes from entanglement spectra:} The existence of symmetry protected gapless edge modes is one of the defining properties of the SPT phase.  
The BIQH state has two counter-propagating edge modes, which can be identified as a charge mode that carries charge with no (pseudo) spin, and  a (pseudo) spin mode that carries (pseudo) spin with no charge. 
Thus, as long as the $U(1)$ symmetry (charge conservation) is preserved, backscattering between the two edge modes is prohibited \cite{Senthil2013}. 

The BIQH state can be described by an Abelian Chern-Simons theory with the $K$-matrix $K=\left( \begin{matrix} 0 & 1 \\ 1 &0 \end{matrix} \right)$ \cite{Lu2012, Senthil2013}. 
Thus the behavior of its edge modes is similar to the ones of FQH states \cite{Wen_book}, whose effective Lagrangian is:
\begin{equation}
\mathcal L=-\frac{1}{4\pi} (K_{\alpha\beta} \partial_t \phi_\alpha \partial_x \phi_\beta+V_{\alpha\beta} \partial_x \phi_\alpha \partial_x \phi_\beta),
\end{equation}
where $\alpha, \beta=A, B$ and $1/2\pi \partial_x \phi_\alpha$ gives the density of the corresponding species of bosons, and $V_{\alpha\beta}$ is the velocity matrix. 
To diagonalize the above Lagrangian, we introduce the charge and (pseudo) spin modes $\phi_{c(s)}=(\phi_a\pm \phi_b)/\sqrt 2$. 
\begin{table}
\setlength{\tabcolsep}{0.19cm}
\renewcommand{\arraystretch}{1.4}
\caption{\label{table:counting_small} The degeneracies of energy levels of the entanglement Hamiltonian (edge mode) in two different charge sectors $\Delta N_a+\Delta N_b=0$ and $\Delta N_a+\Delta N_b=1$.}
\begin{tabular}{ccccccccc}
\hline \hline
  Levels       		 			&  	mode        	& $k_y$ 	   	&  $0$-sector  & $1$-sector             \\ \hline
 Ground state             			&      -   			& 0 			&        1      & 2     \\ \hline
\multirow{2}{*}
{\tabincell{c}{1st excited state}}		&    charge       		&$2\pi/L_y$ 	&         1    & 2    \\ 
                						& 	spin			&$-2\pi/L_y$& 3   & 2 \\ \hline
\multirow{3}{*}
{\tabincell{c}{2nd excited state}}		&    charge   		&$4\pi/L_y$ & 2  &4  \\ 
							&	mixed		&$0$	  & 3  & 2 \\
							&	spin			&$-4\pi/L_y$&4&6\\ \hline
\multirow{4}{*}
{\tabincell{c}{3rd excited state}}		&    charge   		&$6\pi/L_y$ & 3 & 6 \\
							&	mixed		&$2\pi/L_y$  & 6  & 4 \\
							& 	mixed		&$-2\pi/L_y$ &4  & 6 \\
							&	spin			&$-6\pi/L_y$ &7  & 8 \\ \hline\hline
						
\end{tabular}
\end{table}

We can now obtain the edge Hamiltonian and the corresponding momentum operator \cite{Furukawa2013}:
\begin{equation}
H=\frac{2\pi}{L_y}(v_c L_0^c+v_s L_0^s), \quad\quad P=\frac{2\pi}{L_y}( L_0^c- L_0^s), \label{eq:two_modes}
\end{equation}
with
\begin{equation}
L_0^{c(s)}=\frac{(\Delta N_a\pm \Delta N_b)^2}{4}+\sum_{m=1}^{\infty} m n_m^{c(s)}. \label{eq:counting}
\end{equation}
Here, $L_y$ is the length of the 1D edge; $\Delta N_{a(b)}$ is the change in the particle number of $a(b)$ boson relative to the ground state; $\{ n_m^{c(s)}\}$ is the set of non-negative integers describing oscillator modes. These oscillator modes exhibit the well-know $1$, $1$, $2$, $3$, $\dots$ degeneracy pattern \cite{Wen_book}.

As compared to  FQH states with only one chiral mode, Eq.~(\ref{eq:two_modes}) shows two counter propagating modes and thus the BIQH is a non-chiral phase.
Here we briefly mention how to understand the spectrum of the edge modes of the BIQH state.  
Basically, we expect $v_c \sim v_s$ (when $\lambda=0$, this is exact  due to the symmetry of the Hamiltonian), thus $L_0^c+L_0^s$ determines which level (ground state, 1st excited state, ...) of the edge Hamiltonian  the mode belongs to. 
For example  (when $\Delta N_a+\Delta N_b=0$), $L_0^c+L_0^s=0$ gives the ground state and we have $L_0^c+L_0^s \neq 0$ for the excited states.
Specifically if $L_0^s=0$, the edge mode is a charge mode with positive momentum; if $L_0^c=0$, it is a (pseudo) spin mode with negative momentum; if $L_0^c, L_0^s\neq 0$, it is a mixed mode with both charge and (pseudo) spin. 
Once the value of $L_0^c, L_0^s$ is fixed, we have the freedom to choose different combination of $\Delta N_{a(b)}$ and $\{ n_m^{c(s)}\}$  to realize the demanded $L_0^c$ and $L_0^s$. 
Consequently each mode has certain degeneracies for a given energy and momentum.
From Eq. (\ref{eq:two_modes}) and Eq. (\ref{eq:counting}), the degeneracies of each mode can be calculated straightforwardly, and is shown in Table \ref{table:counting_small} (see the supplementary materials for more details).

Numerically, we can use the entanglement spectra as a probe of the edge modes \cite{Li2008}. 
The numerical results from the DMRG simulation are shown in Fig. \ref{fig:ES}. 
We have plotted two different cases that correspond to the $U(1)$ charge sector $\Delta N_a+\Delta N_b=0$ and $\Delta N_a+\Delta N_b=1$. 
The two counter propagating edge modes, and the mixed modes are clearly seen. 
The counting in each sector from our numerics agrees well with the theoretical expectation in Table \ref{table:counting_small}. 
If a finite $\lambda$ is turned on so that the (pseudo) spin is no longer a good quantum number, the (pseudo) spin mode is still robust due to the chiral (anomalous) implementation of  $U(1)$ symmetry \cite{Chen2011,Levin2012,Lu2012, Senthil2013}.
This is also true in our numerical results, we find that the entanglement spectra of finite $\lambda$  remain almost the same as $\lambda=0$ (Fig. \ref{fig:ES}).%

\begin{figure}
\includegraphics[width=0.49\textwidth]{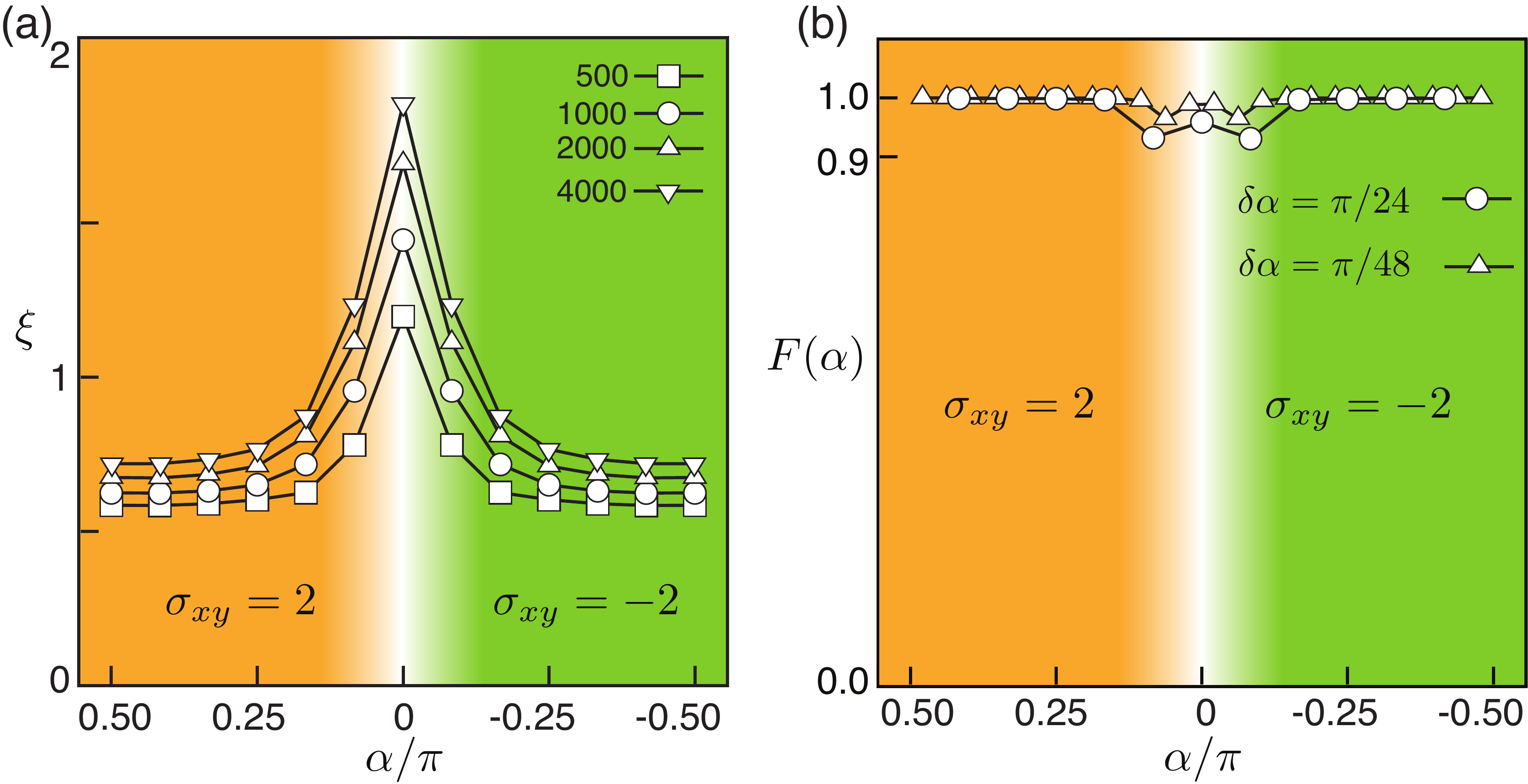} 
\caption{\label{fig:phase_diagram}(color online) Numerical data on the phase transition, the cylinder's width is $W_y=8$ sites. (a) Correlation length $\xi$ in charge sector $\delta n=2$ with different number of states kept in DMRG simulation. (b) Fidelity (wave-function overlap), $F(\alpha)=|\langle \psi(\alpha-\delta \alpha) | \psi(\alpha+\delta \alpha)\rangle|$.}
\end{figure}

\textbf{Phase transition:} We study the phase transition between the two different BIQH states with varying flux $\alpha$.
Interestingly, our numerical results suggest that an infinitesimal flux $\alpha$  drives the system into a BIQH states.
This result is reasonable since any finite flux $\alpha$  breaks the time-reversal symmetry, hence favors a BIQH phase.
Similar physics occurs in Haldane's honeycomb model \cite{Haldane1988}, where an infinitesimal flux stabilizes a Chern insulator. 
To investigate the properties of the phase transition, we calculate the correlation lengths and fidelity (wave-function overlap) for different fluxes $\alpha$.
The correlation length is calculated using the transfer matrix defined in the infinite DMRG's algorithm, which determines the largest  correlation length  \cite{McCulloch2008} (see the supplementary materials).
When the system is deep in the BIQH phase ($\alpha\sim\pm \pi/2$), the DMRG simulations are fully converges and the correlation lengths (Fig. \ref{fig:phase_diagram}(a)) are very small. 
In contrast, near the critical point ($\alpha=0$), the correlation length diverges as the bond dimension is increased. 
This behavior indicates that the system is near a critical point (or phase), where the energy gap is small.
In addition we measure  the fidelity (wave-function overlap) $F(\alpha)$ (Fig. \ref{fig:phase_diagram}(b)), $F(\alpha)=|\langle \psi(\alpha-\delta \alpha) | \psi(\alpha+\delta \alpha)\rangle|$, as $\alpha$ evolves. 
The fidelity ($>0.9$) is very close to $1$, thus we can exclude a strong first order phase transition (which involves level crossing).
However, the DMRG simulations are difficult near the critical point as the bond dimensions diverges, hence a more detailed study (particularly taking account of finite size effects) is required to make definite statements about the nature of the transition.

In general, it is possible that the transition between the two opposite BIQH is a direct continuous phase transition. 
However, it is important to know whether such direct continuous phase transition requires some additional lattice symmetry \cite{Lu2014}, such as the inversion symmetry between $A$, $B$ sublattice of the honeycomb lattice. 
A detailed study of the phase transition and the development of a critical theory require more refined numerical simulations which are left for a future work.  

\textbf{Conclusion and Outlook: }We have introduced a microscopic lattice model that realizes a bosonic integer quantum Hall (BIQH) phase which represents a $U(1)$ symmetry protected topological phase.
The hardcore bosonic model is defined on a honeycomb lattice with correlated hopping and background flux. 
Using  DMRG simulation on an infinite cylinder, we find that the ground state shows two characteristic properties of BIQH: (i) quantized Hall conductance $\sigma_{xy}=\pm 2$ and (ii) two counter propagating gapless edge modes. 
This model  would be an analog of the Chern insulator \cite{Haldane1988}, it is of sufficient simplicity to be relevant for optical lattice experiments \cite{Jotzu2014,Aidelsburger2015} and might furthermore serve as a guide to find new physical realization of topological ordered states \cite{He_prepare}. 

Although we have proved numerically that our model has a BIQH ground state, it is very demanding to develop a microscopic theory for it. 
The correlated hopping term plays  an essential role for the emergence of the BIQH phase, and it couples the bosons in an interesting way (which naively is a mutual flux attachement \cite{Senthil2013}):  if there is a boson at site $k$, the hopping from site $i$ to $j$ is the same as normal hopping;  if there is no boson at site $k$, the boson that hops from site $i$ to $j$ gains a negative sign. 
It is interesting to investigate whether the correlated hopping term is also useful to engineer other interacting SPT phases. 
A possible extension is to realize a BIQH state with a larger Hall conductance (such as $\sigma_{xy}=4$) by assigning a more complicated background flux in our model. 
Another intriguing  direction is to study the topological phase transition between two BIQH phases or  one BIQH phase to a topologically trivial phase. 
For example, by adding some repulsion terms, it is possible to drive the system into a topological trivial Mott insulating phase; hence one might be able to obtain a deconfined phase transition between the BIQH and a topological trivial Mott insulator in our model \cite{Grover2013,Lu2014}.

\emph{Acknowledgement.}---We thank Yohei Fuji for the collaboration on the related project. This work was supported by the Deutsche Forschungsgemeinschaft (DFG) through the collaborative research centre SFB 1143.

\emph{Note added.}---After our preprint occurred on the arXiv, a related work \cite{Sterdyniak2015} appeared introducing an alternative lattice model to stabilize a bosonic integer quantum Hall phase.

\pagebreak
\vspace{5cm}
\widetext

\begin{center}
{\bf \large Supplementary Materials of ``Bosonic integer quantum Hall effect in an interacting lattice model"}
\end{center}

\section{Algorithm }
\begin{figure}[h]
\centering
\includegraphics[width=0.7\textwidth]{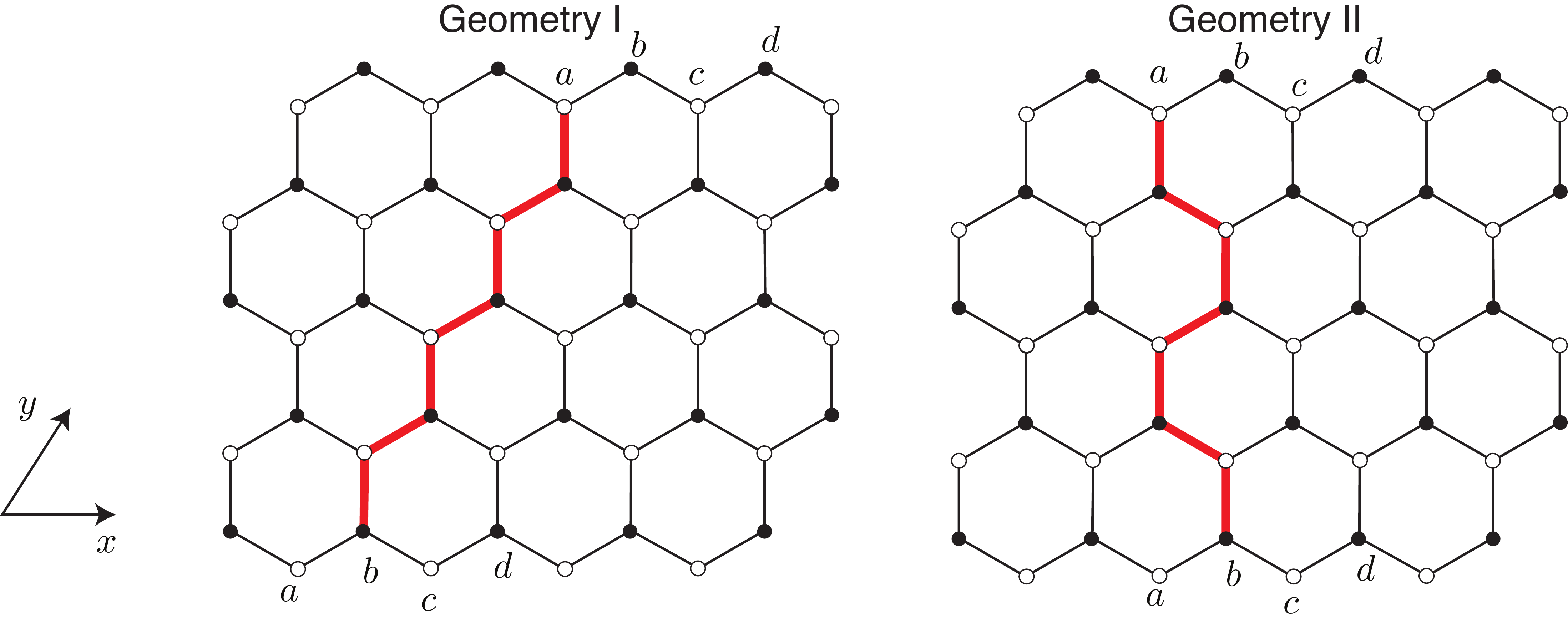} \caption{\label{fig:geo_sup} Two cylinder geometries that we use in the numerical simulations. The red thick line is the one-dimensional path we choose in the DMRG simulation.}
\end{figure}

In our DMRG simulations, we use two ways to wrap honeycomb lattice on a cylinder, as shown in Fig. \ref{fig:geo_sup}. We choose open boundary conditions along the $x$ direction, and periodic boundary conditions along the $y$ direction (with identical sites labeled by the same letter).

\begin{figure}[h]
\centering
\includegraphics[width=0.88\textwidth]{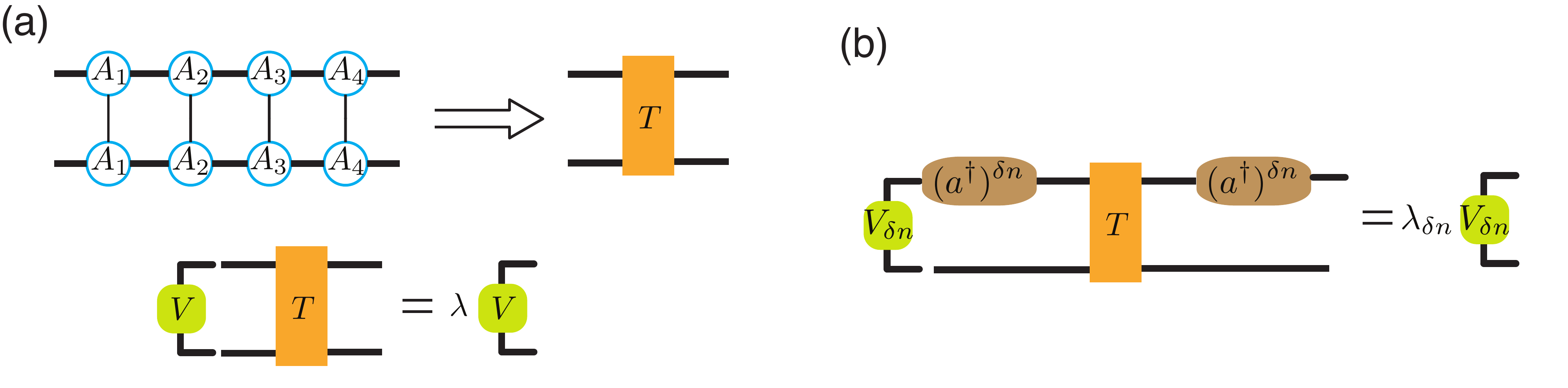} \caption{\label{fig:transfer}  (a) Definition of transfer matrix and correlation length $\xi_{\delta n=0}$ in charge-$0$ sector. (b) Definition of correlation length in charge-$\delta n$ sector.}
\end{figure}

To calculate the correlation length, we define the transfer matrix  $T$ as in  Fig. \ref{fig:transfer} (a). Then the correlation length of the charge-$0$ sector is defined by the first and second largest eigenvalue $\lambda_{1,2}$ (here  $\lambda_1$ is normalized to 1) of the transfer matrix $T$,
\begin{equation}
\xi_{\delta n=0}=-1/\ln \lambda_2.
\end{equation}
The correlation in the charge-$\delta n$ sector is defined similarly, with the quantum number (charge) shifted by $\delta n$ in one leg of the transfer matrix, as shown in Fig. \ref{fig:transfer}(b), to calculate the corresponding dominant eigenvalue $\lambda_{\delta n}$. Then the corresponding correlation length is
\begin{equation}
\xi_{\delta n}=-1/\ln \lambda_{\delta n}.
\end{equation}

The correlation length $\xi_{\delta n}$ determines the largest correlation length in the infinite cylinder \cite{McCulloch2008}, for all possible operator-operator correlation function:
\begin{equation}
\langle \hat O_1^\dag(0) \hat O_2(r)\rangle \sim e^{-r/\xi}, \quad\quad \xi \le \xi_{\delta n}.
\end{equation}
Where the operator  $\hat O_1^\dag$ creates charge $\delta n$, and $\hat O_2$ annihilates charge $\delta n$. For example $b^\dag_l a^\dag_j b_k$ creates $1$ charge, $b^\dag_l a^\dag_j b_k a_i$ creates $0$ charge.

Therefore, instead of calculating various correlation functions, one can simply calculate this single quantity $\xi_{\delta n}$ to know the length scale of the largest possible correlation length.

\section{Convergence of DMRG}

\begin{figure}[h]
\centering
\includegraphics[width=0.75\textwidth]{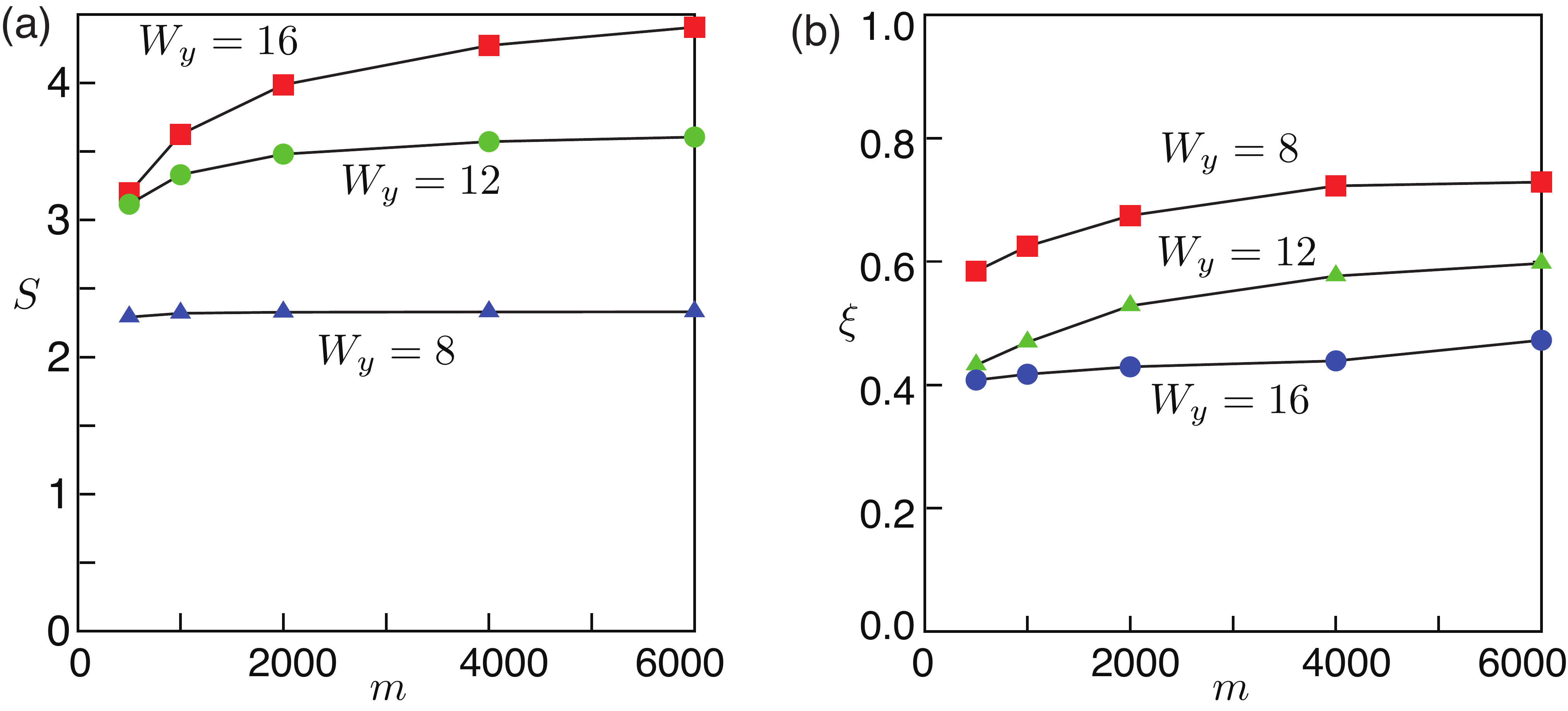} \caption{\label{fig:corr_sys} (a) Entropy versus different number of states kept. (b) Correlation length versus different number of states kept. Here we show $W_y=8, 12, 16$.}
\end{figure}

To see the whether our DMRG simulations converge, we plot the entropy $S$ and correlation length $\xi$ under different system size, in Fig. \ref{fig:corr_sys}, for the system deep in the BIQH phase ($\alpha=\pi/2$). From the entropy, we find that for $W_y=8, 12$, simulations have fully converged as we keep 6000 states. Specifically, the entropy difference as we keep 4000 and 6000 states is, 
\begin{equation}
\frac{S(6000)-S(4000)}{S(6000)}\approx 0.0002, ~~ W_y=8; ~~~~ \frac{S(6000)-S(4000)}{S(6000)}\approx 0.009, ~~ W_y=12.
\end{equation}
For $W_y=16$, the simulation is not fully converged, where $(S(6000)-S(4000))/S(6000)\approx 0.03$. But it's already good enough for us to extract many physical properties.

From the correlation length, we can see that ground state is gapped, since the correlation length is small for all the system sizes, and it decreases as the system size increases. 

\section{The counting of the edge modes in the BIQH state}
The edge Hamiltonian and the corresponding momentum operator of a BIQH state is \cite{Furukawa2013}:
\begin{equation}
H=\frac{2\pi}{L_y}(v_c L_0^c+v_s L_0^s), \quad\quad P=\frac{2\pi}{L_y}( L_0^c- L_0^s), \label{eq:two_modes}
\end{equation}
with
\begin{equation}
L_0^{c(s)}=\frac{(\Delta N_a\pm \Delta N_b)^2}{4}+\sum_{m=1}^{\infty} m n_m^{c(s)}. \label{eq:counting}
\end{equation}

Here we  show how to obtain the  degeneracies of each edge mode, according to Eq. \ref{eq:two_modes}, \ref{eq:counting}.
For example, the spin mode of the 3rd excited state (in $\Delta N_a+ \Delta N_b=0$ sector) corresponds to $L_0^c=0$ and $L_0^s=3$. Then we have two possibilities, $\Delta N_a=\Delta N_b=0$ or $\Delta N_a-\Delta N_b=\pm 2$:
\begin{enumerate}
\item $\Delta N_a=\Delta N_b=0$, thus $\sum m n_m^s=3$. Correspondingly,
\begin{itemize}
\item $n_1^s=3$, and $n_m^s=0$ if $m\ne 1$
\item  $n_1^s=1$, $n_2^s=1$, and  $n_m^s=0$ if $m\ne 1,2$
\item  $n_3^s=1$, and  $n_m^s=0$ if $m\ne 3$
\end{itemize}
 This gives 3-fold degeneracy. This  is why  a single chiral mode (free bosonic oscillator) shows the well known $1$, $1$, $2$, $3$, $\cdots$ degeneracy pattern.
\item $\Delta N_a-\Delta N_b=\pm 2$, thus $\sum m n_m^s=2$. Similar as above, the bosonic oscillator  contributes a $2$ fold degeneracy; together with the $2$-fold degeneracy from $\Delta N_{a(b)}$, we have a $2\times 2=4$ fold degeneracy.
\end{enumerate}
Thus the spin mode of the 3rd excited state has $7$ fold degeneracy. Similarly, we obtain the results for the sector $\Delta N_a+ \Delta N_b=0$ shown in Table \ref{table:counting}, and for the sector $\Delta N_a+ \Delta N_b=1$ shown in Table \ref{table:counting2}. The counting rule for the edge mode is very complicated (although fundamentally straightforward), and our numerical data agrees with the counting rule.

\begin{table}[h]
\setlength{\tabcolsep}{0.2cm}
\renewcommand{\arraystretch}{1.4}
\caption{\label{table:counting} The energy levels of entanglement Hamiltonian (edge mode): $\Delta N_a+\Delta N_b=0$.}
\begin{tabular}{ccccccccc}
\hline \hline
  Levels       		 			&  	mode        	& $k_y$ 	& $L_0^c$, $L_0^s$ 		& $\Delta N_a$, $\Delta N_b$ 		& $\{ n_m^c\} $ 	& $\{ n_m^s\} $     	& Degeneracy                   \\ \hline
 Ground state             			&      -   			& 0 		& $L_0^c=L_0^s=0$ 		&  $\Delta N_a=\Delta N_b=0$ 		& $ n_m^c=0$ 		& $n_m^s =0$  		&        1           \\ \hline
\multirow{3}{*}
{\tabincell{c}{1st excited state \\ 
$L_0^c+L_0^s=1$}} 			&    charge       		&$2\pi/L_y$ 	& $L^c_0=1$, $L^s_0=0$  & $\Delta N_a=\Delta N_b=0$   		& $\sum m n_m^c=1$ &	$n_m^s =0$  	&         1        \\ \cline{2-8} 
                					& \mr{2}{*}{spin} 	& \mr{2}{*}{$-2\pi/L_y$}&  \mr{2}{*}{$L^c_0=0$, $L^s_0=1$}  & $\Delta N_a-\Delta N_b=\pm 2$  &  $ n_m^c=0$ & $n_m^s =0$  & \mr{2}{*}{3} \\    
			         		&            			&  				&  							& $\Delta N_a=\Delta N_b= 0$  &  $ n_m^c=0$ & $\sum m n_m^s=1$   &  &                   \\   \hline
\multirow{5}{*}
{\tabincell{c}{2nd excited state \\ 
$L_0^c+L_0^s=2$}} 	                &    charge               & $4\pi/L_y$  	& $L^c_0=2$, $L^s_0=0$  & $\Delta N_a=\Delta N_b=0$  & $\sum m n_m^c=2$  &$n_m^s =0$  & 2              \\ \cline{2-8}
                 				& \mr{2}{*}{mixed}   & \mr{2}{*}{$0$}  & \mr{2}{*}{$L^c_0=L^s_0=1$} &   $\Delta N_a=\Delta N_b=0$  & $\sum m n_m^c=1$  &$\sum m n_m^s=1$  &  \mr{2}{*}{3}                   \\
                                       		&                   		&  			&  						&  $\Delta N_a-\Delta N_b=\pm 2$ &$\sum m n_m^c=1$ &	$n_m^s =0$  &     \\    \cline{2-8}       
						& \mr{2}{*}{spin}   & \mr{2}{*}{$-4\pi/L_y$}  & \mr{2}{*}{$L^c_0=0$, $L^s_0=2$} &   $\Delta N_a=\Delta N_b=0$  & $ n_m^c=0$  &$\sum m n_m^s=2$  &  \mr{2}{*}{4}                   \\
                                       		&                   		&  			&  						&  $\Delta N_a-\Delta N_b=\pm 2$ &$n_m^c=0$	 &	$\sum m n_m^s=1$  &     \\     \hline
\multirow{7}{*}
{\tabincell{c}{3rd excited state \\ 
$L_0^c+L_0^s=3$}} 	                &    charge               & $6\pi/L_y$  	& $L^c_0=3$, $L^s_0=0$  & $\Delta N_a=\Delta N_b=0$  & $\sum m n_m^c=3$  &$n_m^s =0$  & 3             \\ \cline{2-8}
                 				& \mr{2}{*}{mixed}   & \mr{2}{*}{$2\pi/L_y$}  & \mr{2}{*}{$L^c_0=2$, $L^s_0=1$} &   $\Delta N_a=\Delta N_b=0$  & $\sum m n_m^c=2$  &$\sum m n_m^s=1$  &  \mr{2}{*}{6}                   \\
                                       		&                   		&  			&  						&  $\Delta N_a-\Delta N_b=\pm 2$ &$\sum m n_m^c=2$ &	$n_m^s =0$  &     \\    \cline{2-8}       
						& \mr{2}{*}{mixed}   & \mr{2}{*}{$-2\pi/L_y$}  & \mr{2}{*}{$L^c_0=1$, $L^s_0=2$} &   $\Delta N_a=\Delta N_b=0$  & $\sum m n_m^c=1$  &$\sum m n_m^s=2$  &  \mr{2}{*}{4}                   \\
                                       		&                   		&  			&  						&  $\Delta N_a-\Delta N_b=\pm 2$ &$\sum m n_m^c=1$ &	$\sum m n_m^s=1$  &     \\    \cline{2-8}       
		
						& \mr{2}{*}{spin}   & \mr{2}{*}{$-6\pi/L_y$}  & \mr{2}{*}{$L^c_0=0$, $L^s_0=3$} &   $\Delta N_a=\Delta N_b=0$  & $ n_m^c=0$  &$\sum m n_m^s=3$  &  \mr{2}{*}{7}                   \\
                                       		&                   		&  			&  						&  $\Delta N_a-\Delta N_b=\pm 2$ &$n_m^c=0$	 &	$\sum m n_m^s=2$  &     \\     \hline
\end{tabular}
\end{table}

\begin{table}[h]
\setlength{\tabcolsep}{0.19cm}
\renewcommand{\arraystretch}{1.4}
\caption{\label{table:counting2} The energy levels of entanglement Hamiltonian (edge mode): $\Delta N_a+\Delta N_b=1$.}
\begin{tabular}{cccccccc} 
\hline \hline
  Levels       		 			&  	mode        	& $k_y$ 	& $L_0^c$, $L_0^s$ 		& $\Delta N_a$, $\Delta N_b$ 		& $\{ n_m^c\} $ 	& $\{ n_m^s\} $     	& Degeneracy                   \\ \hline
 Ground state             			&      -   			& 0 		& $L_0^c=L_0^s=1/4$ 		&  $\Delta N_a-\Delta N_b=\pm 1$ 		& $ n_m^c=0$ 		& $n_m^s =0$  		&        2           \\ \hline
\multirow{2}{*}
{\tabincell{c}{1st excited state \\ 
$L_0^c+L_0^s=3/2$}} 			&    charge       		&$2\pi/L_y$ 	& $L^c_0=5/4$, $L^s_0=1/4$  & $\Delta N_a-\Delta N_b=\pm 1$   		& $\sum m n_m^c=1$ &	$n_m^s =0$  	&         2        \\ \cline{2-8} 
              						& spin 			&$-2\pi/L_y$& $L^c_0=1/4$, $L^s_0=5/4$  & $\Delta N_a-\Delta N_b=\pm 1$  		&  $ n_m^c=0$ 			& $\sum m n_m^s=1$  & 2  \\ \hline	
\multirow{4}{*}
{\tabincell{c}{2nd excited state \\ 
$L_0^c+L_0^s=5/2$}} 	                &    charge               & $4\pi/L_y$  	& $L^c_0=9/4$, $L^s_0=1/4$  & $\Delta N_a-\Delta N_b=\pm 1$  & $\sum m n_m^c=2$  &$n_m^s =0$  & 4              \\ \cline{2-8}
                 				& \mr{1}{*}{mixed}   & \mr{1}{*}{$0$}  & \mr{1}{*}{$L^c_0=L^s_0=5/4$} &   $\Delta N_a-\Delta N_b=\pm 1$  & $\sum m n_m^c=1$  &$\sum m n_m^s=1$  &  \mr{1}{*}{2}                   \\  \cline{2-8}       
						& \mr{2}{*}{spin}   & \mr{2}{*}{$-4\pi/L_y$}  & \mr{2}{*}{$L^c_0=1/4$, $L^s_0=9/4$} &   $\Delta N_a-\Delta N_b=\pm 1$  & $ n_m^c=0$  &$\sum m n_m^s=2$  &  \mr{2}{*}{6}                   \\
                                       		&                   		&  			&  						&  $\Delta N_a-\Delta N_b=\pm 3$ &$n_m^c=0$	 &	$n_m^s=0$  &     \\     \hline
\multirow{6}{*}
{\tabincell{c}{3rd excited state \\ 
$L_0^c+L_0^s=7/2$}} 	                &    charge               & $6\pi/L_y$  	& $L^c_0=13/4$, $L^s_0=1/4$  & $\Delta N_a-\Delta N_b=\pm1$  & $\sum m n_m^c=3$  &$n_m^s =0$  & 6             \\ \cline{2-8}
                 				& \mr{1}{*}{mixed}   & \mr{1}{*}{$2\pi/L_y$}  & \mr{1}{*}{$L^c_0=9/4$, $L^s_0=5/4$} &   $\Delta N_a-\Delta N_b=\pm 1$  & $\sum m n_m^c=2$  &$\sum m n_m^s=1$  &  \mr{1}{*}{4}   \\ \cline{2-8}       
						& \mr{2}{*}{mixed}   & \mr{2}{*}{$-2\pi/L_y$}  & \mr{2}{*}{$L^c_0=5/4$, $L^s_0=9/4$} &   $\Delta N_a-\Delta N_b=\pm 1$  & $\sum m n_m^c=1$  &$\sum m n_m^s=2$  &  \mr{2}{*}{6}                   \\
                                       		&                   		&  			&  						&  $\Delta N_a-\Delta N_b=\pm 3$ &$\sum m n_m^c=1$ &	$n_m^s=0$  &     \\    \cline{2-8}       		
						& \mr{2}{*}{spin}   & \mr{2}{*}{$-6\pi/L_y$}  & \mr{2}{*}{$L^c_0=1/4$, $L^s_0=13/4$} &   $\Delta N_a-\Delta N_b=\pm 1$  & $ n_m^c=0$  &$\sum m n_m^s=3$  &  \mr{2}{*}{8}                   \\
                                  		&                   		&  			&  						&  $\Delta N_a-\Delta N_b=\pm 3$ &$n_m^c=0$	 &	$\sum m n_m^s=1$  &  
   \\     \hline
\end{tabular}
\end{table}

\end{document}